# Composition design of refractory compositionally complex alloys using machine learning models

Tao Liang, Eric A. Lass, Haochen Zhu, Carla Joyce C. Nocheseda, Philip D. Rack, Stephen Puplampu, Dayakar Penumadu, and Haixuan Xu*
*Email: xhx@utk.edu

ABSTRACT

Refractory compositionally complex alloys (RCCAs) are considered the next generation of high-temperature materials with great application prospects due to their high performance at elevated temperatures. However, their high-dimensional composition spaces are too large to explore by traditional density functional theory or experimental means, making new RCCA discovery slow and cumbersome. This work has addressed these challenges with an integrated composition design framework that can efficiently and exhaustively explore the relationship between the compositions and two fundamental aspects: 1) the phase stability, including the target body-centered cubic (BCC) phase and its competing phases (hexagonal closed-pack (HCP) structures, Laves and B2 intermetallic phases), and 2) the mechanical properties. This framework is demonstrated with RCCAs within nine refractory metals (Ti, V, Cr, Zr, Nb, Mo, Hf, Ta, and W). Theory-guided machine learning (ML) models were employed to find the composition-mechanical property relationship of RCCAs, where the established theory is used to supplement the yield strength data at ultra-high temperature, and a forward sequential feature selection (SFS) is used to determine feature selection. The resulting ML model for temperature-dependent yield strength was found to have an R_squared value of 0.98 over the entire temperature range (from 0 to 2000 K). The impact of each constituent element on the six key properties is evaluated. The addition of Nb tends to stabilize the BCC phase and the addition of Ti improves the ductility of RCCAs. Combined with all methods involved in this framework, the on-demand designer allows the alloy designers to have all properties for any RCCA compositions and narrow down the composition space by applying custom screening criteria. The output from the predictor and screener provides valuable guidance for our experimental study of RCCAs and accelerates the pace of materials discovery.

Keywords: RCCAs, composition design, machine learning

INTRODUCTION

Compositionally complex alloys (CCAs) [1-7], which are also referred to as high entropy alloys (HEAs), multi-principal-element alloys (MPEAs), concentrated solid solution alloys (CSSAs), or complex concentrated alloys (CCAs), are solid solutions (SS) consisting of multiple principal elements. The inclusion of multiple principal elements has vastly increased the dimensionality of the composition space of CCAs. Consider, if every 10% change in concentration is considered a distinct CCA base, then it is possible to form up to 592 billion unique compositions using 3 to 6 principal elements selected from 75 common metals [8]. This enormously high dimensionality in composition space offers potential materials solutions to meet the industrial needs, and on the other hand, it also demands new alloy design strategies that can navigate the relationship between composition and thermodynamic and mechanical properties of interest in a very efficient way.

One fundamental question for the composition design is to address the phase present at the thermodynamic equilibrium, which is governed by the relative Gibbs free energy of all possible phases ($\Delta G^{phase}$) of a given CCA composition. If only the enthalpy of formation ($\Delta H^{phase}$) is available, the $\Delta G^{phase}$ can be calculated by $\Delta H^{phase} - T\Delta S^{phase}$, where T is the temperature, and $\Delta S^{phase}$ is the entropy of the phase, in which the configurational part can be estimated with an ideal solution assumption. In principle, the $\Delta G^{phase}$ and/or $\Delta H^{phase}$ are accessible by the high-fidelity computational methods such as density functional theory (DFT) based methods and calculation of phase diagrams (CalPHAD) methods. DFT methods provide unbiased enthalpy of formation ($\Delta H$) of CCAs. However, a full DFT treatment of CCAs often involves sufficiently large supercells to accommodate the macroscopic random alloys, which makes it computationally expensive to explore the composition space of CCAs completely. Another approach, DFT with the coherent potential approximation (CPA), substitutes random alloys with a perfectly ordered effective medium. This single-site approximation has dramatically reduced computational effort compared with the supercell method within standard DFT [8, 9]. However, CPA method fails to capture the local lattice distortion (LLD), often leading to an overestimation of the $\Delta H$ [10-12]. The CalPHAD method [13, 14] is a semi-empirical approach that uses thermodynamic theories and experimental databases to calculate phase diagrams. While CalPHAD is often considered the most reliable computational method due to its agreement with experimental data, it is often limited by the scope of the experimental databases it relies on. Since different computational methods have their strengths and limitations [10-12], it is essential to combine and leverage the strengths of different methods to enhance the predictive capabilities for phase stability of CCAs. While ML models to predict thermodynamic properties exist [10, 15-19], the purpose of this study is not to create an ML model to predict thermodynamic properties. Rather, we use these high-fidelity computational methods to directly evaluate the phase stability across all compositions.

It is impractical to perform large-scale simulations of the mechanical properties of CCAs with computational methods alone. Instead, analytical models [20, 21] or ML models [22-26] are often employed to find the correlation between the composition and mechanical properties of CCAs. For example, Maresca et al. [20] developed a dislocation-based analytical model to estimate the temperature-dependent yield strength of alloys. They generated a database of yield strength for over $10^7$ CCA compositions and identified more than $10^6$ possible alloy compositions with high yield strengths for future exploration [27]. However, these analytical models for mechanical properties are oversimplified. ML-based models are better suited for capturing the nonlinear responses and the underlying physics for mechanical properties [23-25]. For example, Giles et al [24] have developed temperature-dependent yield strength ML models for RCCAs using 314 experimental measurements. The mean absolute error (MAE) for their gradient boosting regression model was 118 MPa. By contrast, the analytical model developed by Maresca et al. [20] has an MAE of 683 MPa for the same dataset. Despite these advancements, ML-based frameworks informed by experimental data remain an underutilized approach for discovering CCA compositions with desirable mechanical properties. This is primarily due to the limited availability of experimental data of CCAs. As the experimental database continues to expand, it is worthwhile to redevelop the ML models with the updated database to improve the accuracy in predicting the mechanical properties of CCAs.

In this study, we have demonstrated our composition design strategy to quickly explore the complete composition space of the body centered cubic (BCC) single-phased refractory CCAs (RCCAs) within nine selected refractory metals. An overview of the composition design process used in this study is shown in Fig. 1. Nine refractory metals comprised of elements in Groups 4 (Ti, Zr, and Hf with hexagonal closed-pack (HCP) structures), 5 (V, Nb, and Ta with BCC structures), and 6 (Cr, Mo, and W with BCC structures) of the periodic table are under consideration, which is labelled the "Element space" in Fig. 1. Assuming every 10% change in concentration gives a distinct refractory CCAs (RCCAs) base, our composition space (CS) contains 466 multicomponent (ternary to novenary) systems and 43425 RCCAs bases/compositions. Since the composition space contains all possible compositions, it is labelled as "Exhaustive composition space". Our composition design strategy is focused on providing a systematic approach that integrated the computational methods for phase stability and ML models for mechanical properties to efficiently quantify two fundamental aspects of all RCCA bases: 1) phase stability including the competing phases such as single-phase HCP structures, Laves, and B2 intermetallics (IM), and 2) mechanical properties including the temperature-dependent yield strength, Vicker's hardness and compressive ductility.

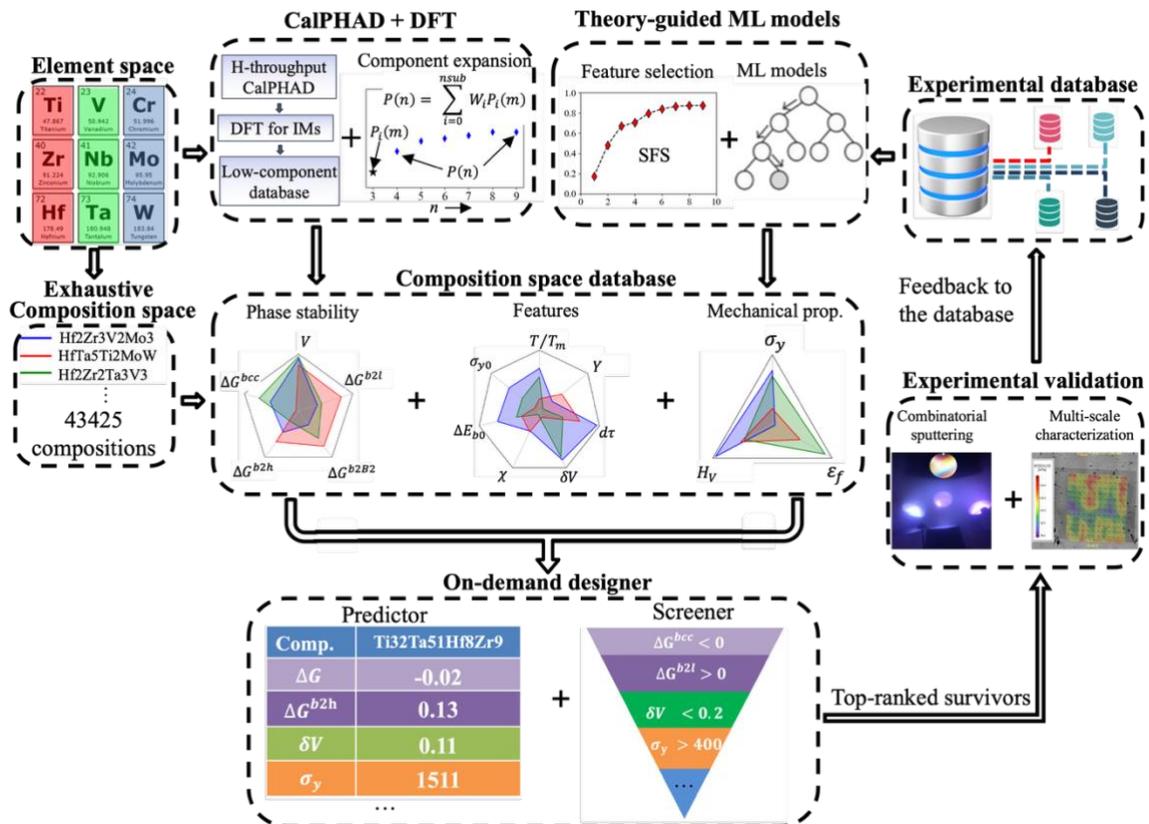

Fig. 1. Overview of composition design strategies. "Element space": A list of nine refractory metals in this work, where elements are colored by columns in periodic table. "CalPHAD+DFT": The high-throughput CalPHAD (H-throughput CalPHAD) is mainly used to get the thermodynamic properties of single-phase SS of low-component (binary and ternary) RCCAs. DFT is used to compute the enthalpy of formation of binary Laves and B2 intermetallic (IM) phases. Once the databases for low-component systems are completed, the thermodynamic properties of a n-component (n>2) system ($P(n)$) of single-phase SS as well as IM phases can be obtained by component expansion (labelled as

"Component expansion"). "Theory-guided ML models": The features for a machine learning (ML) model are selected by feature selection procedure, in which the number of features is determined by the forward sequential feature selection (SFS) method. "Experimental database": Database collection and recompilation from available datasets, and feedback from experimental validation. "Exhaustive composition space": If every 10% change in concentration gives a distinct RCCAs base, our exhaustive composition space (CS) has 43425 distinct RCCA (from ternary to novenary systems) compositions. Three random compositions are chosen to show the content in composition space (CS) database. "Composition space database": CS database for our exhaustive composition space. Phase stability category contains atomic volume ($V$) and thermodynamic quantities, which include Gibbs energies of formation for bcc structure ($\Delta G^{bcc}$) and phase transformation energies from bcc to hcp ($\Delta G^{b2h}$), laves ($\Delta G^{b2l}$), and B2 ($\Delta G^{b2B2}$) phases. Features category contains all feature values (see feature list in Supplementary Information for details). Mechanical properties are predicted values from ML models for yield strength ($\sigma_y$), Vickers hardness ($H_V$), and compression at fracture ($\varepsilon_f$). "Experimental validation": Combinatorial thin film sputtering and multi-scale mechanical-property characterization. "On-demand designer": Predictor is used to get all quantities in CS database for a given off-grid composition and Screener is used to down select compositions with custom screening conditions.

The thermodynamic properties of these 43425 RCCAs are obtained by a combinatorial approach that integrates DFT methods and CalPHAD. In particular, the atomic volume ($V$) and Gibbs free energy of mixing of single-phase BCC ($\Delta G^{bcc}$) and HCP ($\Delta G^{hcp}$) SS of the low component (binary and ternary) systems are obtained by high-throughput CalPHAD calculations ("h-throughput CalPHAD") using its integrated Python software package [28]. The energy difference between BCC and HCP phases ($\Delta G^{b2h} = \Delta G^{hcp} - \Delta G^{bcc}$) is then calculated to evaluate the phase order between these two single-phase SS. The thermodynamic properties of high-component (quaternary to novenary) systems, which are not provided in CalPHAD database, is obtained by the "component expansion" (CE), i.e., extrapolating the thermodynamic properties of high-component systems from the lower-component systems. Since the CalPHAD database has very limited data for two major competing IM phases (the Laves and B2), we used the standard DFT method to evaluate the phase stability problem. The $\Delta H$ of all possible binary IM systems are computed using DFT, and then the site compositions of a given RCCA on sublattices in these two IM phases is determined by the composition most likely to have the lowest $\Delta H$ of IM. For Laves IM, we proposed one third of larger atoms on A sublattice and two thirds of smaller atoms on B sublattice, since larger atoms are always energetically favorable on A sublattice in Laves phase for all binary Laves systems. We have introduced the sequential pair deployment (SPD) approach to determine the site compositions for B2 phase. After the site compositions are determined, the $\Delta H$ of a given RCCA in these two IM phases can be expanded from the constituent binary systems, which is termed "CE for IM phases" (see METHODS section for details) in this study. The energy difference between the BCC and Laves ($\Delta G^{b2l}$) and B2 ($\Delta G^{b2B2}$) phases are used to evaluate the phase stability of RCCAs, where the configurational entropy of mixing ($\Delta S$) is approximated by an ideal solution.

The second part of this work is the ML loop (right side of Fig. 1), which combines "theory-guided ML models" and our combinatorial experimental synthesis and characterization techniques. Here, we use experimental data to train the ML models and then to predict the mechanical properties of our exhaustive CS compositions. Three mechanical properties, including temperature-dependent yield strength ($\sigma_y$), Vickers hardness ($H_V$), and compression at fracture ($\varepsilon_f$) are considered for machine learning (ML) models. The experimental database is often small and has limited coverage of the chosen features. The theory-guided data augmentation, as well as feature selection, is proposed to improve the model transferability. The final selection of features of each ML model is determined by a forward sequential feature selection ("SFS") method [29, 30] to find the most informative feature set. Once the features for a given ML model are determined, the final ML model is trained with the gradient boosting regression (GBR) framework [31].

The on-demand designer is realized by integrating a predictor and a screener. The predictor, which combines the methods in both the phase stability determination and the three ML models for mechanical properties, can predict the thermodynamic and mechanical properties for any composition. The screener is to efficiently narrow down the composition space by applying custom screening criteria. The selected compositions from the predictor and screener will guide our combinatorial experimental study of RCCAs. The experimental data will add to the existing database to further improve the ML models.

The paper is organized as follows. We briefly discussed the phase stability study and elaborated the development of the temperature-dependent yield strength ML model. The methods involved in the phase stability study of this work such as CE and SPD are all presented in METHODS section. The feature calculation and selection that are associated with development of ML models are also found in METHODS. The Vicker's hardness and compression at fracture ($\varepsilon_f$) models are provided in Supplementary Information.

## THERMODYNAMIC PROPERTIES FOR PHASE STABILITY

Three computational methods, including the standard DFT method using Vienna Ab initio Simulation Package (VASP) [32, 33], the Exact Muffin-Tin Orbital with Coherent Potential Approximation (CPA) DFT method [8, 34], and CalPHAD method using ThermoCalc (TC) version 2023b [28] were employed to get the thermodynamic properties in this work. The computation details are provided in Supplementary Information. Since TC method is considered as the most reliable approach, we selected TC method to get the thermodynamic data for single-phase SS of RCCAs. As for IM phases, only 10 out of 72 possible binary Laves compositions are available in TC database and no binary B2 data is available in TC database. We therefore selected the standard DFT method to calculate the $\Delta H$ for all binary IM phases. Though DFT-CPA method overestimates the $\Delta H$ of a given composition, it can be used to estimate the properties that are relevant to the energy difference, such as the elastic properties with the same level of accuracy of the standard DFT method. In this work, the CPA method is only used to evaluate the accuracy of the elastic constants from the rule of mixture (ROM).

### Single-phase Solutions

Since the ground state of these 9 refractory metals is either HCP or BCC phase, other single-phase solutions such as face centered cubic (FCC) were skipped in this study. The atomic volume ($V$), thermodynamic quantities $\Delta H$, $\Delta S$ and $\Delta G$ of BCC and HCP phases for all elemental, binary, and ternary compositions are obtained from throughout TC calculations with its integrated "TC_python" software package [28]. The TC database provides full coverage of binary compositions of interest. However, 277 out of 3024 ternary compositions were absent for the BCC phase, and 483 out of 3024 ternary compositions were absent for the HCP phase. The missing data for ternary compositions were obtained from the CE (see METHODS section for details). This completed database of ternary compositions has then served as the base database, from which the thermodynamic properties of high-component (from quaternary to novenary) compositions can be quickly obtained by CE. The energy difference between BCC to HCP phases ($\Delta G^{b2h}$) of all RCCAs, which is defined as $\Delta G^{hcp}$-$\Delta G^{bcc}$, is then calculated.

### Laves and B2 intermetallic phases

Laves with an AB$_2$ formula and B2 with AB formula are two commonly observed IM phases in these RCCAs, where A and B represent their sublattice sites, respectively. Unlike the single-phase solutions, there are numerous ways to deploy atoms on A and B sublattices for a given RCCA composition. Thus, the first step for these two IM phases is to find the compositions on A and B sublattice sites (site compositions). Due to data shortage in TC database, we used the DFT method to calculate the $\Delta H$ of all binary Laves and B2 phases. For Laves binary phase, it is always energetic favorable to have larger radius element on A site and smaller radius element on B site. Thus, for a given RCCA composition, the site composition deployment with one third of atoms with larger radii to A site and two thirds of atoms with smaller radii B site is most likely having the lowest the formation of energy of Laves phase ($\Delta H^{laves}$). The site compositions for B2 phase are determined by the SPD (see METHODS section for details). Once the site compositions of a given RCCA composition are determined, the $\Delta G^{laves}$ and $\Delta G^{B2}$ of the RCCA is estimated by "CE for IM phases" (see METHODS section for details). Likewise, the energy difference between BCC to Laves phase ($\Delta G^{b2l} = \Delta G^{laves} - \Delta G^{bcc}$) and B2 phases ($\Delta G^{b2B2} = \Delta G^{B2} - \Delta G^{bcc}$) were computed to evaluate phase order of the competing phases. Since thermodynamic properties from DFT and TC might have significant difference, the $\Delta H^{bcc}$ inside $\Delta G^{bcc}$ of a given RCCA composition are taken DFT values and its component expansion technique provided and developed by Bokas and et al [35].

**ML MODELS FOR MECHANICAL PROPERTIES**

Three mechanical properties, which include the temperature dependent compressive yield strength ($\sigma_{y0}$), Vickers hardness ($H_V$), and the compression at fracture ($\varepsilon_f$), were considered in this study. A gradient boosting regression (GBR) model for each of mechanical properties was developed based on the experimental database in literatures [23, 36, 37]. The inputs for ML models were more than 30 features that were encoded from their compositions (see "Feature calculation" in METHODS section). To improve the transferability of the ML models that were trained from the limited experimental database and then used to predict the mechanical properties of the CS of interest, we proposed a theory-guided feature selection strategy (see "Feature selection" in METHODS section) to find the best feature set for a given ML model. Once the features for a given mechanical property were decided, we have developed a series of ML models and chose the best ML model as the final model to predict the corresponding mechanic property for the CS. Below, we have elaborated the development of ML models for the temperature dependent yield strength ($\sigma_y$). The Vickers hardness model and the compression at fracture ($\varepsilon_f$) model were provided in Supplementary Information.

**Experimental data preparation for yield strength model**

The experimental temperature dependent yield strength data were collected from the databases compiled by Borg et al [37] and by Couzinie et al. [36]. The original data contained both compression and tension yield strength data for a variety of compositions, phases, processing conditions, and grain sizes. Only compression data were used in this work as the tension data are insignificant in number. Since this work is primarily focused on composition design for BCC single phase solid solutions, we firstly excluded the materials without BCC phase presented in their microstructure. Secondly, the processing conditions (e.g. synthesized method and annealing temperature) and grain size are neglected, and the recorded data were the mean value of various processing conditions and grain sizes. After recompiling the original database, 530 data points with 228 unique compositions were left, in which only 82 compositions had their yield strength measured at least two temperature levels and only 6 data points have the measurement temperature above 1500 K. To improve predictions at high temperatures, these 82 compositions with multiple temperatures were subject to data augmentation, where the thermally-activated temperature-dependent yield strength model proposed by Maresca and Curtin (MC model) [20] (Eq. (1)) was used to extrapolate yield strength at temperatures up to 2000 K.

$$\sigma_y(T, \dot{\varepsilon}) = \begin{cases} \sigma_{y0}\left[1 - (\frac{kT}{\Delta E_{b0}} \ln \frac{\dot{\varepsilon}_0}{\dot{\varepsilon}})^{2/3}\right] &, \sigma_y \geq 0.5\sigma_{y0} \\ \sigma_{y0} \exp\left(-\frac{1}{0.55}\frac{kT}{\Delta E_{b0}} \ln \frac{\dot{\varepsilon}_0}{\dot{\varepsilon}}\right) &, \sigma_y < 0.5\sigma_{y0} \end{cases} \quad (1)$$

In Eq. (1), $\sigma_{y0}$ is the zero-temperature flow stress, $k$ is the Boltzmann constant, $\dot{\varepsilon}_0$ is the reference strain rate (set to $10^4$ s$^{-1}$), $\dot{\varepsilon}$ is the strain rate (set to $10^{-3}$ s$^{-1}$), and $\Delta E_{b0}$ is the dislocation energy barrier. For data augmentation, we set threshold temperature to 800K, above which $\sigma_y$ is assumed to be less than $0.5\sigma_{y0}$. 61 out of 82 compositions have at least two data points with temperatures above 800K, then two unknown variables in Eq. (1) ($\sigma_{y0}$ and $\Delta E_{b0}$) are obtained from a curve fitting to lower part of Eq. (1). The proposed extrapolation temperatures were set to four levels of 1100, 1400, 1700 and 2000 K, in which the starting extrapolation temperature has to be 150 K larger than the highest measured temperature and the highest extrapolation temperature has to be smaller than 0.95 of its melting temperature ($T_m$), where $T_m$ is estimated from a rule of mixtures (ROM). Lastly, the extrapolated yield strength has to be larger than 5 MPa. After the data augmentation, the final dataset used for model training and validation had 628 temperature-dependent yield strength values with 295 data points measured at low temperatures (temperature < 800 K) and 333 data points at high temperatures (temperature >= 800 K).

**GBR model for yield strength model**
The feature selection for yield strength model follows the feature selection procedure in Methods Section. The final selected features are $T/T_m$ (the ratio between the measured temperature and melting point), $\sigma_{y0}$, $\Delta E_{b0}$, $\chi$ (electronegativity), $dV$ (atomic volume distortion), $d\tau$ (shear modulus distortion), and $Y$ (Young's modulus). As input features, $\sigma_{y0}$ and $\Delta E_{b0}$ are calculated by Eq. (18) and (19) in Ref. [20] with suggested values for BCC RCCAs. Once the selected features are determined, 628 experimental yield strength datapoints were used to train and validate the gradient boosting regression model using the library in *scikit-learn* package [38], where minimum samples of a leaf node are set to six to prevent from the overfitting. We have repeated this model training/validating process 20 times and taken the GBR model with the highest coefficient of determine ($R^2$) as our final model.

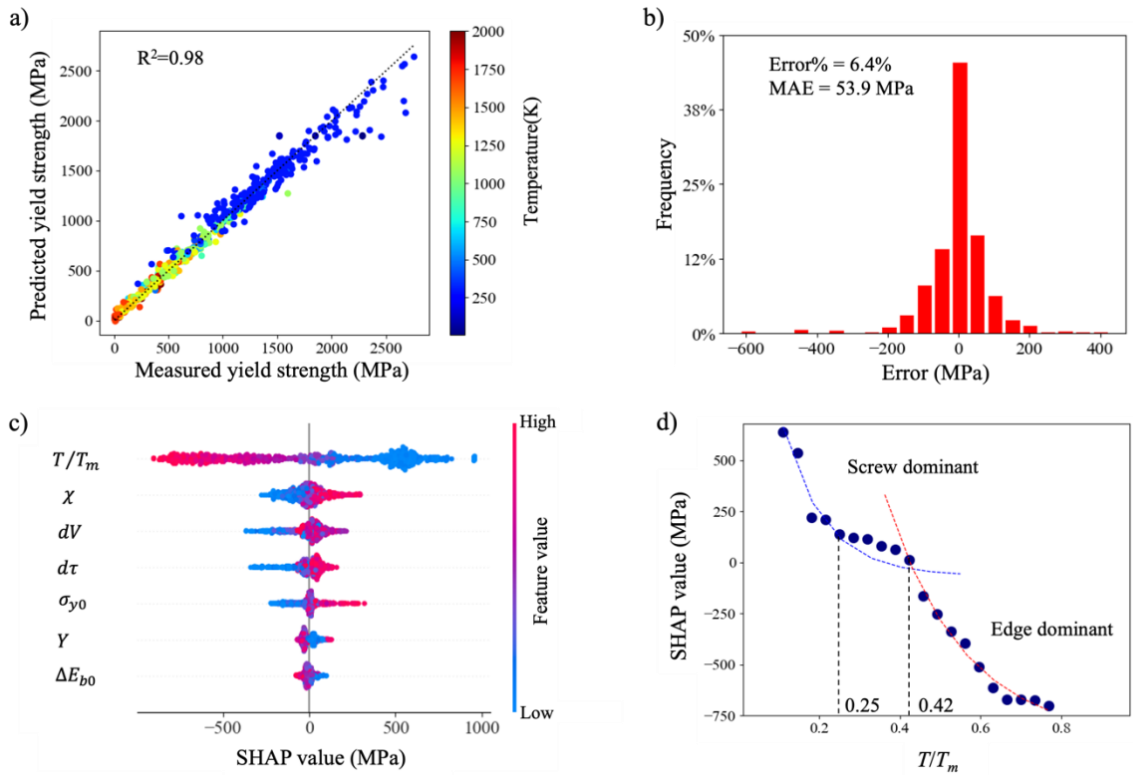

Fig. 2. Yield strength model performance, error analysis, SHAP value of all features, SHAP value of $T/T_m$ of yield strength model. a) The parity plot comparing predicted yield strength with measured yield strength, where all data are colored by measured temperatures; b) error (predicted yield strength – measured yield strength) distribution analysis, where MAE is the mean absolute error; c) SHAP values of all model features, where features are presented in a descending order of their importance; and d) SHAP value of $T/T_m$, which indicates a transition from screw- to edge-dominant dislocation strengthening from low $T/T_m$ (<0.25) to high $T/T_m$ (>0.42) ranges.

Fig. 2a shows the parity plot between measured and predicted yield strength data of the best GBR model, where all data are colored by measured temperatures. The model was found to have a very good quantitative agreement (R_squared value = 0.98) with the experimental database over the entire temperature range (from

0 to 2000 K). The errors (predicted $\sigma_y$ – measured $\sigma_y$) show a normal distribution with a mean value near zero (Fig. 2b). The mean absolute error (MAE) is about 54.2 MPa, which corresponds to a percentage error of 6.4%. The well performance of our model is indicative of the importance of these features. The key feature $T/T_m$ is used to capture temperature dependence and the Young's moduli ($Y$) is the slope of stress-strain curve. The $\sigma_{y0}$ and $\Delta E_{b0}$ in MC model reflects the dislocation strengthening, and the volume distortion ($dV$) and shear distortion ($d\tau$) reflects the solid solution hardening induced by local lattice distortion. The electronegative $\chi$, which is highly correlated with Young's modulus ($Y$), is believed that have similar impact with Young's modulus (further discussion is shown in next paragraph).

The Shapley Additive Explanations (SHAP) analysis [39, 40] was applied to quantify the importance of features on the model predictions. SHAP analysis uses a game theoretic approach to compute a SHAP value of a feature or feature value, which determine a collaborative game contributes to the model outputs of the corresponding feature or feature value [39]. Features with positive SHAP values positively impact on the model outputs, while those negative SHAP values have a negative impact. The magnitude is a measure of how strong the effect is. The SHAP values of all features was shown Fig. 2c, where features were sorted by their importance in a descending order. As expected, $T/T_m$ is the dominant feature, with high $T/T_m$ corresponding to a lower yield strength. The $\chi$, $\sigma_{y0}$, $d\tau$, and $dV$ tend to positively contribute to the yield strength. As feature importance/impact might be misleading when some of features are strongly correlated, we have computed R_squared value of a linear regression between all possible pairs of features within 628 datapoints. As shown in Fig. 3S in Supplemental Information, the strongly correlated (R_squared value > 0.7) are $\sigma_{y0}$ and $\Delta E_{b0}$ ($R^2$ = 0.70), $\sigma_{y0}$ and $d\tau$ ($R^2$ = 0.72), and $\chi$ and $Y$ ($R^2$ = 0.82). A SHAP plot of selected features, which are low-ranked in Fig. 2c and low-correlated with each other, is shown in Fig. 3S in Supplemental Information, where only room temperature yield strength is used. Comparing SHAP values in Fig. 2c with Fig. 5S, $\sigma_{y0}$ and $\Delta E_{b0}$ have similar contributions to predicted yield strength. Likewise, $\chi$ and $Y$ contributions to yield strength are similar.

A detail impact of $T/T_m$ on predicted yield strength is plotted in Fig. 6d, where the bin values of $T/T_m$ was taken from the node values of the model estimator and the y-axis is the mean of SHAP values of samples whose $T/T_m$ values are within two neighboring bins. Clearly, the yield stress ($\sigma_y$) response in bcc alloys exhibits three distinct regimes: a strong thermal dependence at low ($T/T_m$ < 0.25) and high ($T/T_m$ > 0.42) temperatures and a relatively insensitive between these two regions. Several studies [41-44] have attributed multiple regimes of temperature-dependent yield strength in bcc alloys to a transition of slip mode from screw to edge dislocation during deformation.

The GBR yield strength model was then used to obtain the yield strength of our exhaustive composition space, where the temperature levels are set to a linear spacing between 0 to 2700 K with a step size of 300 K. If the $T/T_m$ of a given composition is larger than 0.9, the yield strength is set to zero.

**SHAP EVALUATION OF CONSTITUENT ELEMENTS**
The database of the exhaustive CS (43425 data) can be completed by applying the methods introduced in the phase stability and developed ML models. It is worthwhile to quantify the contributions of 9 constituent elements to the key properties. Here, the relationship between molar fraction or concentration of elements and 6 selected properties, which are the Gibbs free energy of mixing ($\Delta G^{bcc}$), energy difference between BCC phase and HCP ($\Delta G^{b2h}$), Laves ($\Delta G^{b2l}$), and B2 ($\Delta G^{b2B2}$) phases, yield strength ($\sigma_y$) at 1500 K, and the compression at fracture ($\varepsilon_f$), were linked by 6 ML models, respectively. The input features for these 6 ML models are the concentration of the 9 elements in the 43425 compositions, and each model was trained randomly selected 39083 data points (90% of 43425 data points). The contribution of 9 constituent elements for each property was evaluated by SHAP analysis on the rest 4342 (10% of 43425 data points) data points. Noted that the CS database contains RCCA composition from ternary to novenary systems with a concentration increment of 0.1, the number frequency was heavily biased on low concentrations as shown in Fig. 3. In specific, there are 19188 data points without a given element (concentration of 0%), 11432 data points having a concentration of 10%, 6427 data points having a concentration of 20%, and only 28 data points having a concentration of 80%, which is the highest possible concentration in CS database. The width of SHAP values of a given element in Fig. 4 reflects the number frequency distribution plot in Fig. 3.

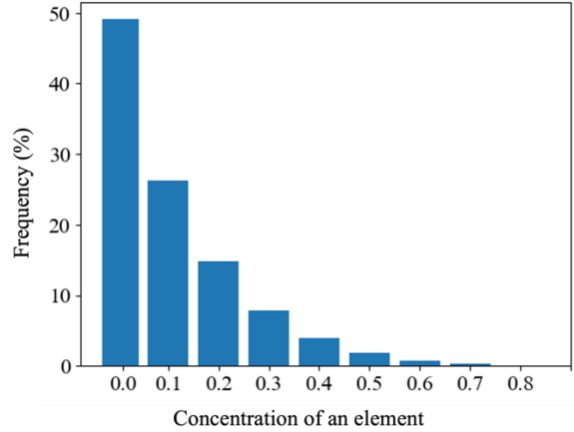

Fig. 3. Distribution of concentration of an element of 43425 compositions in CS

The SHAP analysis of these 6 models on the 4342 data points were plotted in Fig. 4, in which the features were the concentrations of the nine constituent elements and the presence of elements was ordered by the importance of elements. As expected, Group IV elements (Ti, Zr, and Hf), whose ground states are HCP phases, have positive impact on $\Delta G^{bcc}$ and negative impact on $\Delta G^{b2h}$. Group V (V, Nb, and Ta) and Group VI (Cr, Mo and W) elements, whose ground states are BCC phases in general have negative impact on $\Delta G^{bcc}$ and positive impact on $\Delta G^{b2h}$ with the only exception of Cr. Through Cr-X binary phase diagrams in TC database, where X is one of eight elements, Cr either form $XCr_2$ Laves phase or have a large miscibility gap with X elements. Thus, Cr elements have a reserved effected on the phase stability of BCC SS phase. As for Laves phase of $AB_2$ formula, in which the atomic radius ratio between A and B sites are often required to be ranging from 1.13 to 1.24, Cr and V with two smallest radii (1.42 Å and 1.49 Å, see Table 1S in Supplementary Information) and Hf and Zr with two largest radii (1.75 Å and 1.77 Å) tend to favor of the formation of Laves phase for a given RCCA composition. As a result, the $\Delta G^{b2h}$ values decrease as the concentrations of these four elements increase. There is no clear trend on the element group for $\Delta G^{b2h}$ as shown in Fig. 4d. Additions of Mo and Cr tend to form B2 phase. Whereas the addition of Nb tend to stabilize the BCC phase rather than B2 phase. Summarizing the SHAP evaluation on these four thermodynamic quantities, Nb is a good phase stabilizer for the BCC phase. While Cr in general is a detrimental addition to form BCC SS when it is coalesced with other 8 elements, and the concentration of Cr should be low (below 20%) if the target phase is the BCC SS.

The SHAP values of the yield strength at 1500 K (Fig. 4e) and compression at fracture (Fig. 4f) in general reflect the yield strength-ductility paradox in materials design. The Group VI elements (Cr, Mo, and W) with highest yield strength among three groups in general have a positive contribution to the yield strength and negative contribution to the ductility of the RCCAs. As discussed above, the Cr-RCCAs and V-RCCAs tend to form Laves phase. As shown in Fig. 4e, the Laves phase seems to have negative contribution to the yield strength. Since the ratio of temperature and melting temperature ($T_m$) is the dominant feature for yield strength, Ta, whose $T_m$ is 3290 K (see Table 1S in Supplementary Information), has highest positive contribution to the high temperature yield strength in Group V elements. Group IV elements which have lowest yield strength among three group in general decrease the yield strength. Whereas Hf with $T_m$ of 2506 K, which is the highest among Group IV elements, has slightly positive contribution to yield strength at high temperature. Among all 9 constituent elements, only Ti can improve the ductility of RCCAs. The Group V elements and Hf exhibit mixed impacts on the compression at fracture.

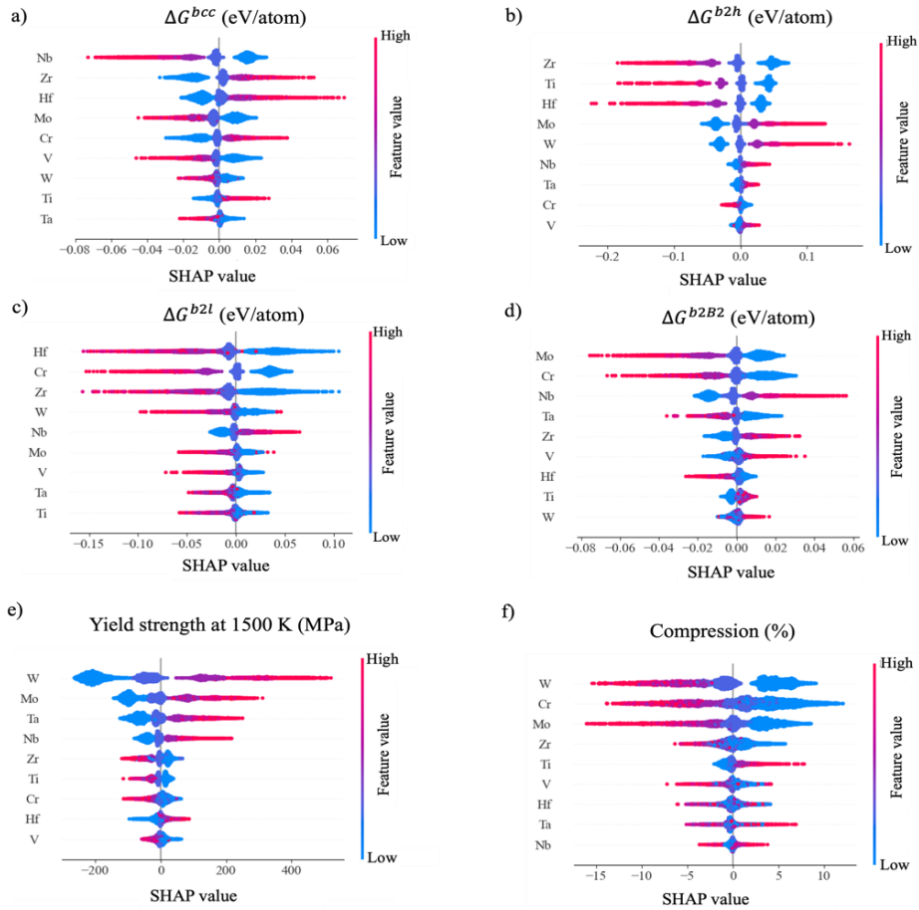

Fig. 4. SHAP values of the nine constituent elements for ML models for selected properties: a) $\Delta G^{bcc}$; b) $\Delta G^{b2h}$; c) $\Delta G^{b2l}$; d) $\Delta G^{b2B2}$; e) Yield strength at 1500 K; and f) Compression at fraction.

The SHAP analysis on 9 constituent elements provides the general trends of the elements' impact on the specific properties, which in return are helpful for composition design of RCCAs.

**PREDICTOR AND SCREENER**

A predictor has combined the methods in phase stability and ML models to get the values of properties, which include thermodynamic data, feature values, and the mechanic properties of a given composition. The screener is to fast narrow down the composition space by applying custom screening criteria. Below, we presented an example of using the predictor and screener to find the most promising equiatomic compositions, which are stable with BCC phase and have high yield strength at high temperature and reasonable compression at room temperature. There are 420 equiatomic RCCAs from ternary to novenary systems. Except quinary systems, the rest equiatomic RCCAs are all off-grid compositions. The database for these 420 equiatomic RCCAs are built by the predictor. As discussed above, the first screening criterion is set to the composition must include Nb and Ti. To have a stable BCC phase, the Gibbs free energy of formation of BCC phase ($\Delta G^{bcc}$) is set to smaller than 0.0 eV/atom, the three phase transformation energies ($\Delta G^{b2h}$, $\Delta G^{b2l}$, and $\Delta G^{b2B2}$) are all set to larger than 0.0 eV/atom. As an empirical criterion to have SS

[45], the volume mismatch is set to smaller than 20%. The $\varepsilon_f$ is set to larger than 15%. The screening for the yield strength at 1500 K is set to larger than 80% of the maximum yield strength. The results after applying the screening conditions are listed in Table 1.

Table 1. List of surviving equiatomic compositions with a custom screener (see manuscript for detail screening criteria). The temperature for the yield strength ($\sigma_y$) is set to 1500 K.

| Composition | $\Delta G^{bcc}$ (eV/atom) | $\Delta G^{b2h}$ (eV/atom) | $\Delta G^{b2l}$ (eV/atom) | $\Delta G^{b2B2}$ (eV/atom) | $\delta V$ (%) | $\varepsilon_f$ (%) | $\sigma_y$ (MPa) |
|---|---|---|---|---|---|---|---|
| TiNbMoHfTa | -0.035 | 0.233 | 0.062 | 0.013 | 10.9 | 23.5 | 621 |
| TiVNbTaW | -0.065 | 0.313 | 0.130 | 0.068 | 11.1 | 20.0 | 633 |
| TiNbMoHfTaW | -0.037 | 0.297 | 0.056 | 0.038 | 11.0 | 17.9 | 678 |
| TiCrNbTaW | -0.005 | 0.296 | 0.059 | 0.003 | 14.7 | 19.3 | 714 |
| TiZrNbMoTaW | -0.028 | 0.273 | 0.072 | 0.044 | 14.4 | 23.4 | 763 |

**APPLICATIONS TO TTHZ ALLOYS**

**CONCLUSIONS**
Navigating interrelationship between composition and material properties is the key mission for materials discovery, which is particularly challenging when confronting with the enormous high dimensionality of the composition space inherited in the CCAs. We have therefore proposed a widely applicable framework for composition design that combines the integrated computational methods and empirical theories of component expansion for the phase stability and machine learning models for mechanical properties. Our method is demonstrated in down selection of the promising compositions among RCCAs from 9 refractory metals – Ti, V, Cr, Zr, Nb, Mo, Hf, Ta, and W.

The calculation of the thermodynamic properties of the exhaustive composition space has integrated the CalPHAD for single-phase SS and DFT method for Laves and B2 phases, the two major competing IM phases in RCCAs of interest. The high throughput CalPHAD calculations are only focused on the binary and ternary compositions and the thermodynamic properties of uncovered compositions are obtained from the component expansion. For the Laves phase, we have assigned one third of atoms with larger radii to A sublattice site and two thirds of atoms with smaller radii B sublattice site for a given RCCA composition. For the B2 superlattice, we have introduced the SPD method to determine compositions on A and B sublattices for a given RCCA composition. After the compositions on sublattices are determined, the Gibbs free energies of formation of the Laves and B2 phases for a given RCCA composition are estimated by the method so called "CE for IM phases".

The mechanic properties are often out of range of computational methods. The machine learning models are used to find the relationship between composition-based features and target mechanic properties. These machine models are trained with sparse experimental data and then is used to predict the corresponding mechanic properties of the composition space of interest. To improve the transferability of machine learning models, we have proposed a theory-guided feature selection strategy for data augmentation and feature selection, in which the feature coverages between experimental database and the composition space of interest are evaluated and the best feature set is determined by a forward sequential feature selection (SFS) method with a repeated $k$-fold cross validation method in the gradient boosting regressor (GBR) models. We have developed three ML models for temperature-dependent yield strength, compression at fracture, and Vickers hardness in this study, respectively.

The impacts of 9 constituent elements on the key properties are investigated by SHAP analysis. Nb is a BCC phase stabilizer and whereas addition of Cr makes the BCC phase less stable. As expected, W and Mo can significantly improve the yield strength, they are detrimental additions for ductility. The addition of Ti can clearly improve the ductility of RCCAs. The rest elements in Groups IV and V have a mixed impact on the compression at fracture.

The methods for the phase stability and three ML models for mechanical properties are linked by the predictor, in which only required input is the composition. The final selection of the promising compositions is screened by applying the well-established criteria for materials design or a set of custom screening functions. These outputs from the predictor and screener provide valuable guidance for our experimental studies.

## METHODS
### Component expansion for single-phase solutions
The component expansion (CE), which is used to extrapolate a property of a high-component composition from its constituent low-component sub-systems, is shown in Eq. (*2*).

$$\begin{cases} P(c_1, c_2, \ldots c_n) = \frac{1}{\Omega} \sum_{isub}^{C_n^m} \Omega_{isub} P(c_{1,isub}, c_{2,isub}, \ldots c_{m,isub}) \\ \Omega_{isub} = \sum_{i=1}^{m} c_i \text{ and } c_{i,isub} = \frac{c_i}{\Omega_{isub}} \\ \Omega = \sum_{isub}^{C_n^m} \Omega_{isub} \end{cases}, \quad (2)$$

In Eq. (*2*), $P$ stands the property of interest, which includes $V$, $\Delta H$, $\Delta S$, $\Delta G$, and $\Delta G^{b2h}$, $c_i$ is the concentration of $i^{th}$ element. The $n$ is the number of components of the *n*-component composition and $m$ ($m < n$) is the number of components of sub-system from which the property was expanded, $C_n^m$ is the number of constituent sub-systems of the *n*-component systems and $c_{i,isub}$ is the normalized concentration of $i^{th}$ element in the *isub* m-component composition. The ratio between $\Omega_{isub}$ and $\Omega$ is the molar fraction of a given sub-composition. Through this equation, the property of a *n*-component composition was a weighted sum of the properties of its constituent *m*-component compositions, where the weight of each *m*-component composition is set to its molar fraction. If $c_{i,isub}$ is off-grid (i.e. cannot divide by 10% exactly), the property was obtained from a N-dimensional interpolation (NDIP) function coded in *scipy* software package [46]. The validation of the component expansion method is provided in Fig. 2S in Supplementary Information.

### Sequential pair deployment for B2 and CE for Laves and B2 intermetallic phases

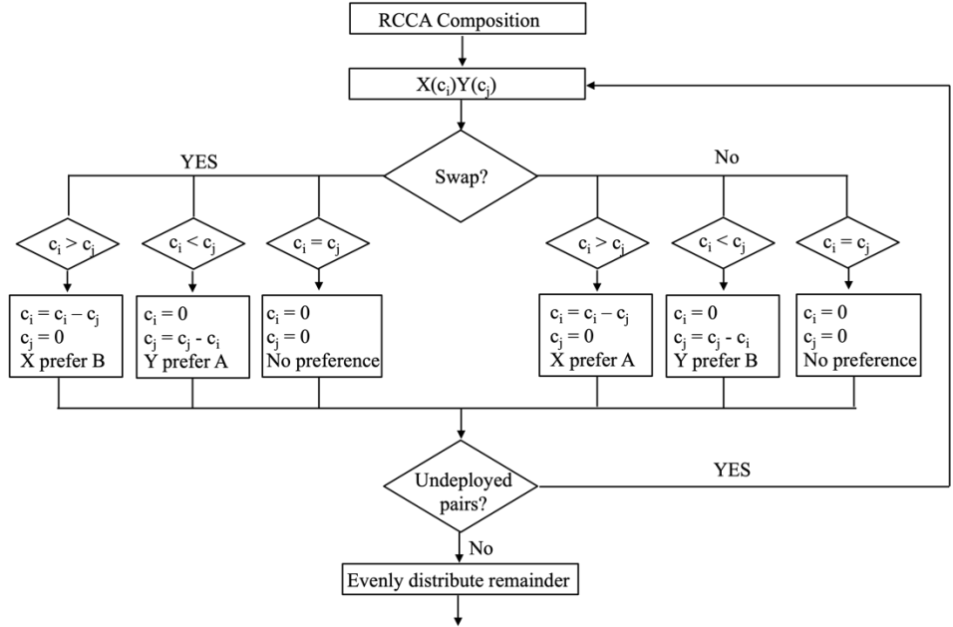

Fig. 5. Flowchart of the sequential pair deployment (SPD) to find the compositions on A and B sublattice sites of the B2 superlattice. $X(c_i)Y(c_j)$ is the next available undeployed XY pair, $c_i$ is the current concentration of X element, and $c_j$ is the current concentration of Y element. If Swap is False, X atoms deploy on A site and Y atoms deploy on B site. Otherwise, X atoms deploy on B site and Y atoms deploy on A site.

The SPD strategy is sketched Fig. 5, where A and B are the sublattice sites of B2 phase. For a given n-component RCCA composition, all possible $C_n^2$ pairs were ascendingly ranked by their $\Delta H^{B2}$ of the pair, where $C_n^2$ is the number of combinations of selecting 2 out of n component. Thus, the lower index in the list of $C_n^2$ pairs indicated a lower value of $\Delta H^{B2}$. As shown in Fig. 5, the SPD strategy started with the XY pair with the lowest $\Delta H^{B2}$, no preference and no swap. The preference in Fig. 2 was used to indicate the site to which the excessive atoms in previous step were deployed. Swap in Fig. 2 is a binary indicator to show whether to swap X and Y atoms on A and B sublattice sites. If Swap is False, X atoms deploy on A site and Y atoms deploy on B site. If there is a preference in previous step (X prefer A, X prefer B, Y prefer A, or Y prefer B), the excessive atoms in previous step were kept on the preferred site. If it results in X on A or Y on B, the Swap in this step is set to False, and otherwise the Swap is set to True. If there is no preference, the radius difference between A and B sublattice sites were computed with and without swap deployment. Since the formation of B2 phase is favorable with smaller atomic radius mismatch, the deployment of the XY pair was assigned to the deployment (swap or no swap deployment) with the smaller radius difference between A and B sites. Depending on current concentrations of X ($c_i$) and Y ($c_j$) types of atoms and the Swap, the type of excessive atoms and its preferred site can be decided. Taking $c_i > c_j$ and Swap is True as an example, the concentration of X type of atoms, which is the excessive type, is $c_i - c_j$, and the Y type of atoms is depleted ($c_j=0$) after deployment. Since Swap is True (X is deployed to B site), the excessive X atoms prefer on B sublattice site (X prefer B in Fig. 2), which was passed to next step to decide the value of Swap. Since Y type of atoms was depleted, the next available pair would be the next highest ranked pair which includes X type of atoms. The SPD strategy recursively deployed the available pair until all the pairs were deployed. If there were remaining atoms after all pairs were deployed, they were evenly distributed on A and B sublattice sites.

$$\begin{cases} \Delta G^{IM} = \Delta H^{IM}(c_1, \dots, c_n) - T\Delta S^{IM}(c_1, \dots, c_n) \\ \Delta H^{IM}(c_1, \dots, c_n) = \sum_i^{n_A} \sum_j^{n_B} \Delta H_{ij}^{IM} c_{i,A} c_{j,B} \\ \Delta S^{IM} = -c_A \sum_i^{n_A} R \ln c_{i,A} - c_B \sum_j^{n_B} R \ln c_{j,B} \end{cases} \quad (3)$$

Once the composition on A and B sites is determined for these two IM phases, the Gibbs free energy of formation of a n-component composition in Laves and B2 IM phases can be estimated Eq. (*3*), where $T$ is the temperature ($T$ = 300 K), $R$ is the gas constant, $n_A$ and $n_B$ are the number of elements on A and B sites, $c_{i,A}$ and $c_{j,B}$ are the normalized concentration of $i^{th}$ element on A site and $j^{th}$ element on B site, and $c_A$ (1/3 for Laves and 1/2 for B2 phases) and $c_B$ (2/3 for laves and 1/2 for B2 phases) are summed concentration of A and B sites, respectively. Adopting the idea of component expansion for single phase, the $\Delta H^{IM}$ of a given RCCA composition is a weighted sum of its sub-binary superlattices, in which the weight of sub-binary IM phase is its molar fraction (the production of $c_{i,A}$ and $c_{j,B}$). The entropy of mixing ($\Delta S^{IM}$) is a weighted sum of the configurational entropy of sublattice sites, where $c_A$ and $c_B$ are the weights on A and sublattice sites. From the number of component point of view, Eq. (*3*) can be considered as a component expansion function for the Gibbs free energy of formation of a n-component composition in Laves and B2 superlattice, which is why it is termed as "CE for IM phases" in this paper.

**Feature calculation**
A variety of composition-based features including the features in other publications [20, 22-24] were calculated for the systems in our exhaustive composition space database and experimental database. These features can be categorized to 1) lattice constant associated quantities such as atomic radius (r), atomic volume (V), atomic volume mismatch ($\delta V$), and volume distortion ($dV$); 2) thermodynamics associated quantities such as entropy ($\Delta S$ ), the enthalpy of formation of bcc phase ($\Delta H^{bcc}$) and phase transformation energy from bcc to hcp phases ($\Delta H^{b2h}$); 3) elastic property associated quantities such as Young's (Y) modulus, shear ($\tau$) modulus, bulk (B) modulus, Poisson's ratio ($\upsilon$) and shear distortion ($d\tau$); 4) constructed quantities such as a dimensionless ratio between entropic and enthalpic contributions ($\Omega^{SH}$) proposed by Giles *et al* [24], the elastic contribution to the enthalpy of formation ($\Delta H^{el}$) using Miedema model [47], and the $\sigma_{y0}$ and $\Delta E_{b0}$ that calculated from the composition using the equations proposed by Maresca and Curtin [20]; and 5) other quantities such as melting point ($T_m$), electronegativity ($\chi$), density ($\rho$), and valence electron concentration (VEC). The elastic properties calculated from DFT are generally comparable to experimental data with an error smaller than 15% for most inorganic materials [48]. However, the EMTO-CPA calculated bulk moduli for Cr and Hf have the errors of 37.34% and 43.11% (Table 1S in Supplemental Information), respectively. By comparing the bulk, Young's and shear moduli for all ternary compositions (Fig. 4S in Supplementary Information), the ROM method has a comparable accuracy with the CPA method. Rather than taking DFT calculated results, the features in Category 3 are thus obtained from the ROM of the experimental data. For compositions within the composition space of interest, i.e. the constituent elements are within the 9 elements of interest, the features in Categories 1 and 2 can be obtained from CE method in Section 1. For the compositions outside the composition space of interest, the features in Categories 1 and 2 were obtained from additional TC calculations for uncovered binary systems. A comprehensive list of the features is provided in Table 2 in Supplementary Information.

**Feature selection**

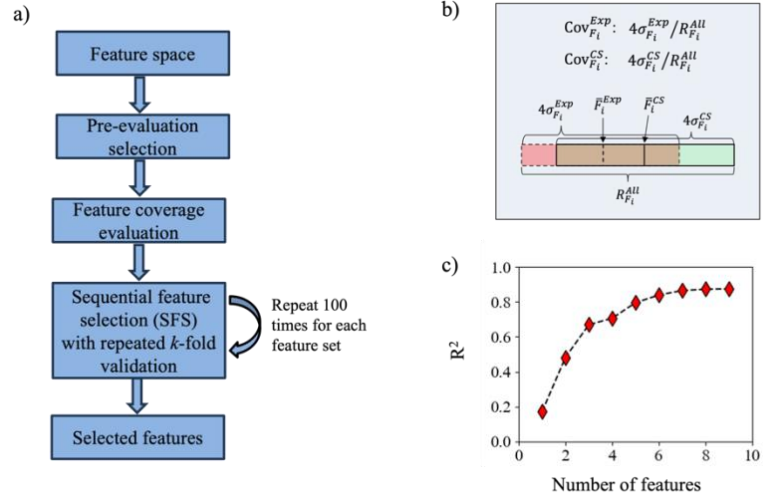

Fig. 6. Theory-guided feature selection strategy. a) Procedure of the theory-guided feature selection; b) Schematic view of the coverage calculation of $i^{th}$ feature ($\text{Cov}_{F_i}$), where superscript "*Exp*" represents the experimental database, "*CS*" represents the exhaustive composition space, $\bar{F}_i$ is the mean value of $i^{th}$ feature, $\sigma_{F_i}$ is the standard deviation of $i^{th}$ feature, and $R_{F_i}^{All}$ is the range of $i^{th}$ feature, which is defined as maximum($\bar{F}_i^{Exp} + 2\sigma_{F_i}^{Exp}, \bar{F}_i^{CS} + 2\sigma_{F_i}^{CS}$) – minimum($\bar{F}_i^{Exp} - 2\sigma_{F_i}^{Exp}$, $\bar{F}_i^{CS} - 2\sigma_{F_i}^{CS}$); c) A typical plot of R_squared value ($R^2$) vs. the number of features in a forwarding sequential feature selection (SFS) process.

Here, we proposed a theory-guided feature selection strategy (Fig. 6) to improve the transferability of ML models. A general procedure of the feature selection is schematically shown in Fig. 6a, where the feature space is listed in Table 2S in Supplementary Information. Prior to the feature coverage evaluation (pre-evaluation selection in Fig. 6), features were first reduced by removing the undefined, the low-variance, and linearly dependent features. In specific, $\Delta G^{b2l}$ and $\Delta G^{b2B2}$ were excluded since they are undefined in some experimental compositions. The Poisson's ratio was excluded since the variance (the ratio between standard deviation and mean value) is low, which is about 7% in CS. Since the atomic volume can be computed from the atomic radius ($V = \frac{4}{3}\pi r^3$), we selected volume-associated features (atomic volume, volume mismatch, and volume distortion) rather than radius-associated features in this study. At the pre-evaluation selection step, some features are forced to include since they might overshadow the importance of the rest features or might be excluded from the following feature evaluation steps. Taking the temperature-dependent yield strength as an example, the ratio between the measured temperature and melting point ($T/T_m$) is the dominant feature [41, 49]. To better indicate yield strength response to the temperature-independent features, the only room-temperature yield strength data are used to evaluate the rest of features. In addition, it is widely accepted that dislocation strengthening mechanism is important for yield strength in crystalline alloys. Thus, the $\sigma_{y0}$ and $\Delta E_{b0}$ in Table 2S in Supplementary Information, which is derived from the energetics of dislocation motion in bcc CCAs [20], are selected at this step and excluded from the following feature evaluation process. The feature coverage evaluation in Fig. 6b is to remove under- or over-covered features from ML models. First, the mean value ($\bar{F}_i$) and the standard deviation ($\sigma_{F_i}$) of a given feature in the experimental database (denoted to "Exp" as a superscript) and in the exhaustive composition space (denoted to "CS" as a superscript) are computed, respectively. Assuming the feature has a Gaussian distribution, 95% datapoints in the experimental database are within the range of $\bar{F}_i^{Exp} \pm 2\sigma_{F_i}^{Exp}$, and the same to the exhaustive composition space. The overall range of the feature ($R_{F_i}^{All}$) in two databases is defined as the difference between maximum value of $\bar{F}_i^{Exp} + 2\sigma_{F_i}^{Exp}$ and $\bar{F}_i^{CS} + 2\sigma_{F_i}^{CS}$ and minimum value of $\bar{F}_i^{Exp} - 2\sigma_{F_i}^{Exp}$ and $\bar{F}_i^{CS} - 2\sigma_{F_i}^{CS}$. Then the feature coverage of a given feature with respected to $R_{F_i}^{All}$ in two databases ($\text{Cov}_{F_i}^{Exp}$ and $\text{Cov}_{F_i}^{CS}$) are computed as Eq. (*4*). If $\text{Cov}_{F_i}^{Exp}$ of $i^{th}$ feature is smaller than a threshold value, this feature is considered as an under-covered feature. If $\text{Cov}_{F_i}^{CS}$ of $i^{th}$ feature is smaller than the threshold value, this feature is considered as an over-covered feature. Both under- and over-covered features are excluded for ML model development. The threshold value set to 0.7 for yield strength and Vickers hardness and set to 0.6 for compression at fracture. Depending on experimental database, several features such as density ($\rho$) and ratio between entropic and enthalpic contributions ($\Omega^{SH}$) are under-covered features and excluded for ML development. The only over-covered feature is valence electron concentration (VEC) as the number of valence electrons of our element space are one of 4, 5, or 6. Though VEC was identified as an important indicator for the ductility in CCAs [50], it was excluded in this study for all models.

$$\text{Cov}_{F_i}^{Exp} = 4\sigma_{F_i}^{Exp}/R_{F_i}^{All}$$
$$, and \qquad (4)$$
$$\text{Cov}_{F_i}^{CS} = 4\sigma_{F_i}^{CS}/R_{F_i}^{All}$$

After the feature coverage evaluation step, the remaining features together with the preselected features are then subjected to a forwarding sequential feature selection (SFS) method [30] for selecting the best feature set with the gradient boosting regression (GBR) model in the *scikit*-learn library [38]. The forwarding SFS method is coupled with a repeated *k*-fold cross-validation technique, where *k* is 5 and repeating *k*-fold 100 times for each feature set. The cross-validation regression coefficient ($R^2$) of each feature set is taken as averaged score of GBR models with the repeated *k*-fold technique. At each stage of forwarding SFS, the model chooses the feature with highest $R^2$ value to add to the feature set of the previous stage. The hyperparameters in GBR model are set as follows: a minimum of 4 samples per leaf node, a learning rate of 0.01 and a maximum of 100 estimators. A typical plot of this forwarding SFS is shown in Fig. 6c. When the number of features is larger than 7, adding new feature has negligible improvement of $R^2$ and, thus, the final selected features are the 7 features with the highest $R^2$ value.

ACKNOWLEDGEMENTS
This research was primarily supported by the National Science Foundation Materials Research Science and Engineering Center program through the UT Knoxville Center for Advanced Materials and Manufacturing (DMR-2309083).

DATA AVAILABILITY
They are available upon the request.

CODE AVAILABILITY
They are available upon the request.

# SUPPLEMENTARY INFORMATION

## COMPUTATIONAL DETAILS
### CalPHAD calculations
The high-throughput CalPHAD calculations were performed with ThermoCalc (TC) 2023b within its database of "High Entropy Alloy v6.1". The molar volume, enthalpy, entropy, and Gibbs free energy of 9 elements, 324 binary, and 3024 ternary compositions in body centered cubic (BCC), hexagonal close packed (HCP), laves, and B2 phases were extracted using its integrated "TC_python" software package. The temperature was set to 300 K throughout TC calculations. The resulting atomic radius ($r$), enthalpy ($\Delta H^{phase}$), entropy ($\Delta S^{phase}$), and Gibbs free energy of formation ($\Delta G^{phase}$) in each of these phases were computed.

### EMTO-CPA calculations
The high-throughput equation of state (EOS) calculations were performed with the Exact Muffin-Tin Orbital (EMTO) with Coherent Potential Approximation (CPA) density functional theory (DFT) method [8, 34], where the generalized gradient approximation (GGA) potentials using the Perdew-Burke-Ernzerhof (PBE)[51] exchange correlation functionals with the full charge density techniques [52] are employed. Except Hf, where the valence electron configuration is set to $4f^{14}5d^26s^2$, the electrons on outermost d and s orbitals of the rest 8 elements are treated as valence electrons. The screened impurity parameter of CPA was set to 0.02 with softcore approximation and the energy convergence was set to $1.0 \times 10^{-6}$ eV/atom. The k-point grid was set to $29 \times 29 \times 29$ for BCC phases and $31 \times 31 \times 21$ for HCP phases, respectively. The energy tolerance was set to $1.0 \times 10^{-6}$ eV. A workflow modified from pyEMTO [53] is developed to automatically submit jobs and analyze EMTO-CPA calculations.

### VASP calculations
Vienna Ab initio Simulation Package (VASP) [32, 33] calculations are performed with the generalized gradient approximation (GGA) potential using the Perdew-Burke-Ernzerhof for solids (PBEsol) exchange correlation functionals [51] [54] [55]. A plane-wave cutoff energy was set to 520 eV. The k-point sampling along each of x, y, and z directions was set to an integer such that the product between the integer and the length of the corresponding lattice vector is larger than 20 Å. The energy convergence was set to $1.0 \times 10^{-6}$ eV and the force convergence was set to -0.01 eV/Å.

## COMPARISON OF COMPUTATIONAL METHODS
### Selected Properties for single elements
Selected Properties for 9 elements were listed in Table 1S. The bulk modulus ($B$) from CPA calculations and atomic volume ($V$) from TC calculations were presented as percentage of error with respect to experimental data [56-59]. The phase transformation energy from BCC to HCP ($\Delta G^{b2h}$), which is defined as the Gibbs free energy difference between HCP and BCC phases ($\Delta G^{hcp} - \Delta G^{bcc}$), from three computational methods were compared. The $\text{Diff}^{b2h}_{\text{CPA2TC}}$ in Table 1S is the $\Delta G^{b2h}$ difference between CPA and TC methods ($\Delta G^{b2h}_{CPA} - \Delta G^{b2h}_{TC}$).

Table 1S Comparison of VASP, CPA, and TC calculations for single elements. The experimental data of melting temperature ($T_m$) were collected from the online database [56]. The atomic radius (r) is computed from the atomic volume ($V$) from the TC method ($V = \frac{4}{3}\pi r^3$). The bulk modulus ($B$) from CPA calculations and atomic radius (r) from TC calculations were presented as percentage of error with respect to experimental data [56-59]. $\Delta G^{b2h}$ is the phase transformation energy from BCC to HCP and $Diff^{b2h}_{CPA2TC}$ is the $\Delta G^{b2h}$ difference between CPA and TC methods ($\Delta G^{b2h}_{CPA} - \Delta G^{b2h}_{TC}$).

| | $T_m$ (K) | $B$ (Error%) | r (Å) | r (Error%) | $\Delta G^{b2h}$ (eV/atom) | | | $\text{Diff}^{b2h}_{\text{CPA2TC}}$ |
|---|---|---|---|---|---|---|---|---|
| Method | Exp. | CPA | TC | TC | VASP | CPA | TC | CPA/TC |

| | | | | | | | |
|---|---|---|---|---|---|---|---|
| Ti | 1941 | 4.19  | 1.62 |  0.05 | -0.10 | -0.09 | -0.05 | -0.04 |
| V  | 2183 | 15.21 | 1.49 | -0.19 |  0.27 |  0.30 |  0.05 |  0.25 |
| Cr | 2180 | 37.34 | 1.42 | -0.90 |  0.42 |  0.46 |  0.04 |  0.42 |
| Zr | 2128 | 4.13  | 1.77 |  0.03 | -0.08 | -0.06 | -0.05 | -0.01 |
| Nb | 2750 | 3.78  | 1.63 |  0.02 |  0.30 |  0.36 |  0.11 |  0.24 |
| Mo | 2896 | -3.93 | 1.55 | -0.08 |  0.44 |  0.51 |  0.12 |  0.39 |
| Hf | 2506 | 43.11 | 1.75 | -0.12 | -0.18 | -0.10 | -0.10 |  0.00 |
| Ta | 3290 | -1.53 | 1.63 |  0.09 |  0.3  |  0.33 |  0.13 |  0.19 |
| W  | 3695 | -3.41 | 1.56 | -0.17 |  0.54 |  0.58 |  0.15 |  0.43 |

**Enthalpy of formation of the BCC phase of binary compositions**

Fig. 7S compares the enthalpy of formation of the BCC phase ($\Delta H^{bcc}$) of binary compositions from CPA, VASP, and TC calculations, in which the VASP datapoints were used the equiatomic binary database in Ref. [35], where the random alloys were generated by the special quasi-random structures (SQS) approach [60].

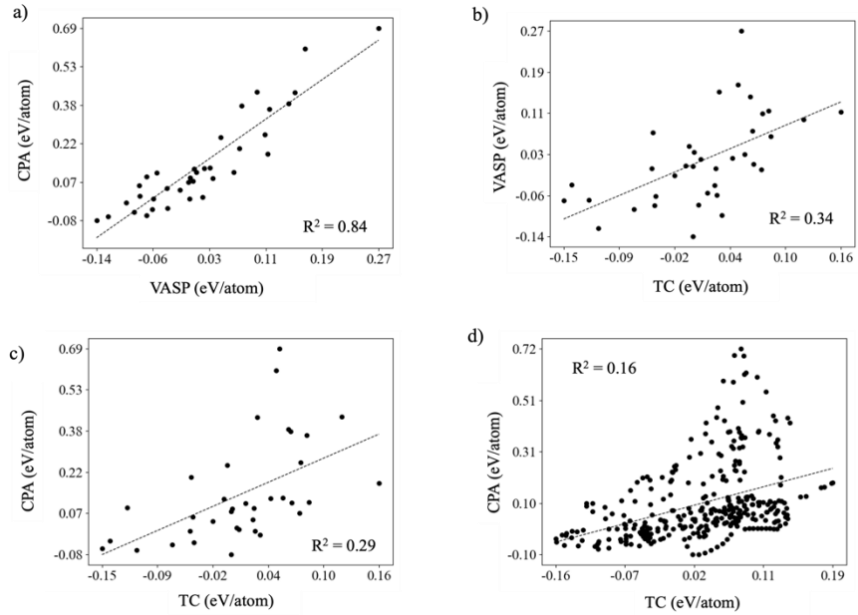

Fig. 7S. Comparison of the enthalpy of formation of the BCC phase ($\Delta H^{bcc}$) of binary compositions from CPA, VASP, and TC methods. Datapoints in plots a), b), and c) are $\Delta H^{bcc}$ for 36 equiatomic binary compositions. The datapoints from VASP calculations were obtained from Ref. [35], where the random alloys were generated by the special quasirandom structures (SQS) approach [60]. Datapoints in plot d) are $\Delta H^{bcc}$ for all 324 binary compositions.

Since CPA method is incapable of capturing the local lattice distortion (LLD) in alloys [10], the CPA significantly overestimates $\Delta H^{bcc}$. As shown in Fig. 7S, the ranges of the $\Delta H^{bcc}$ values were -0.08 to 0.69 eV/atom for CPA calculations and -0.14 to 0.27 eV/atom for VASP calculations. With a linear fitting on CPA and VASP results, the coefficient of determine (R_squared value) is 0.84. It is not surprising that the CPA and VASP predict a similar trend in the enthalpy of formation since they are both DFT methods. However, they both fail to produce the trend of TC results for the enthalpy of formation. The R_squared values for VASP vs. TC and CPA vs. TC were 0.34 and 0.29 for equatomic compositions, respectively. For all 324 binary compositions, the R_squared value of the linear regression fitting between CPA and TC calculations was 0.16. Since the TC method were based on the thermodynamic theories and the experimental data, which was generally considered as a more reliable computational method for thermodynamic quantities, we chose the TC method to acquire atomic volume, the enthalpy of formation, and the Gibbs free energy of formation for single phase SS.

VALIDATION OF COMPONENT EXPANSION

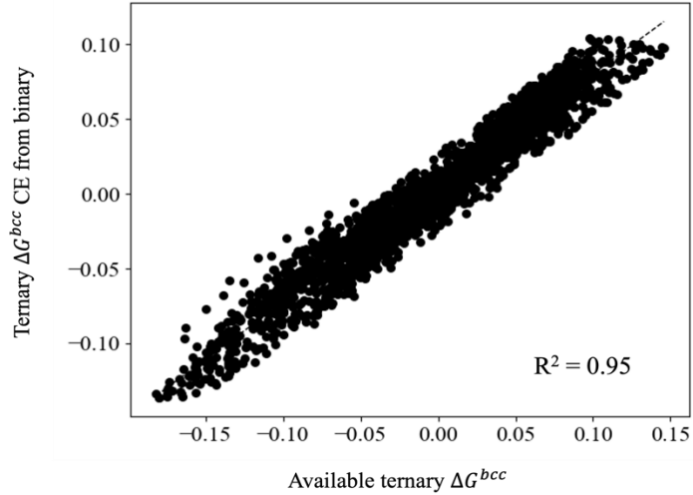

Fig. 8S. Validation of component expansion (CE): The Gibbs free energy of formation for ternary compositions in BCC phase ($\Delta G^{bcc}$). The $\Delta G^{bcc}$ obtained by CE from binary compositions are shown on $Y$ axis and TC results of available ternary compositions (2747 compositions) are shown on $X$ axis.

Using the Gibbs free energy of formation for ternary compositions in BCC phase ($\Delta G^{bcc}$) as an example, the accuracy of component expansion is validated in Fig. 8S, where the results that obtained from CE from binary compositions are shown on $Y$ axis and the original TC results of 2747 available ternary compositions are shown on $X$ axis. The R_square value of the linear regression is 0.95.

MACHINE LEARNING MODELS
**Feature list**
Table 2S Comprehensive list of the compositional-based features in ML models

| Symbol | description | calculation |
|---|---|---|
| a | Lattice constant | From CS database |
| r | Atomic radius | From CS database, $r = \frac{\sqrt{3}}{4}a$ |
| V | Atomic volume | $\frac{4}{3}\pi r^3$ |
| $\delta r$ | Atomic radius mismatch | $\sqrt{\sum_i^n c_i \left(1 - \frac{r_i}{r}\right)^2}$ |
| $dr$ | Atomic radius distortion [24] | $\sum_i^n \frac{9}{8} c_i \sum_{j \neq i}^n c_j \frac{2|r_i - r_j|}{r_i + r_j}$ |
| $\delta V$ | Atomic volume mismatch | $\sqrt{\sum_i^n c_i \left(1 - \frac{V_i}{V}\right)^2}$ |
| $dV$ | Atomic volume distortion [24] | $\sum_i^n \frac{9}{8} c_i \sum_{j \neq i}^n c_j \frac{2|V_i - V_j|}{V_i + V_j}$ |

| | | |
|---|---|---|
| $\Delta S^{phase}$ | Enthalpy of formation of the phase, where $R$ is the gas constant | $-R\sum_{i}^{n} c_i \ln c_i$ |
| $\Delta H^{phase}$ | Enthalpy of formation of the phase | From CS database |
| $\Delta G^{phase}$ | Gibbs free energy of the phase | $\Delta H^{phase} - T\Delta S^{phase}$ |
| $\Delta H^{b2h}$ | Phase transformation enthalpy from BCC to HCP phases | $\Delta H^{hcp} - \Delta H^{bcc}$ |
| $\Delta G^{b2h}$ | Phase transformation Gibbs free energy from BCC to HCP phases | $\Delta G^{hcp} - \Delta G^{bcc}$ |
| $\Delta H^{b2l}$ | Phase transformation enthalpy from BCC to laves phases | $\Delta H^{laves} - \Delta H^{bcc}$ |
| $\Delta G^{b2l}$ | Phase transformation Gibbs free energy from BCC to laves phases | $\Delta G^{laves} - \Delta G^{bcc}$ |
| $\Delta H^{b2B2}$ | Phase transformation enthalpy from BCC to B2 phases | $\Delta H^{B2} - \Delta H^{bcc}$ |
| $\Delta G^{b2B2}$ | Phase transformation Gibbs free energy from BCC to B2 phases | $\Delta G^{B2} - \Delta G^{bcc}$ |
| $Y$ | Young's modulus | $\sum_{i}^{n} c_i Y_i$ |
| $\tau$ | Shear modulus | $\sum_{i}^{n} c_i \tau_i$ |
| $B$ | Bulk modulus | $\sum_{i}^{n} c_i B_i$ |
| $\nu$ | Poisson's ratio | $\sum_{i}^{n} c_i \nu_i$ |
| $d\tau$ | Shear distortion [24] | $\sum_{i}^{n} \frac{9}{8} c_i \sum_{j \neq i}^{n} c_j \frac{2|\tau_i - \tau_j|}{\tau_i + \tau_j}$ |
| $T_m$ | Melting point | $\sum_{i}^{n} c_i T_{m_i}$ |
| $T/T_m$ | Ratio between measured temperature and melting point | $T/T_m$ |
| $\Omega^{SH}$ | Ratio between entropic and enthalpic contributions [24] | $\frac{T_m \Delta S^{bcc}}{\Delta H^{bcc}}$ |
| $\Delta H^{el}$ | Elastic contribution to energy [47], where $C_n^2$ is the number of binary systems from a n-component system | $\sum_{isub}^{C_n^2} \left( c_j \Delta H_{i \; in \; j}^{el} + c_i \Delta H_{j \; in \; i}^{el} \right)$ , and $\Delta H_{i \; in \; j}^{el} = \frac{2B_i \tau_j (V_i - V_j)^2}{4\tau_j V_i + 3B_i V_j}$ |
| $\sigma_{y0}$ | Zero-temperature flow stress [20], where $A_\sigma$ is 0.04, $\alpha$ is 1/12, and $\boldsymbol{b}$ (Burger's vector) is $\frac{\sqrt{3}}{2}a$ | $3.067 A_\sigma \alpha^{-1/3} \tau \left(\frac{1+v}{1-v}\right)^{4/3} \left(\sum_i \frac{c_i(V_i - V)^2}{\boldsymbol{b}^6}\right)^{2/3}$ |
| $\Delta E_{b0}$ | Dislocation energy barrier [20], where $A_E$ is 2.0 | $A_E \alpha^{1/3} \tau \boldsymbol{b}^3 \left(\frac{1+v}{1-v}\right)^{2/3} \left(\sum_i \frac{c_i(V_i - V)^2}{\boldsymbol{b}^6}\right)^{1/3}$ |
| $\chi$ | electronegativity | $\sum_i^n c_i \chi_i$ |
| $VEC$ | Valence electron concentration | $\sum_i^n c_i VEC_i$ |
| $M$ | Atomic mass | $\sum_i^n c_i M_i$ |
| $\rho$ | Density, where $M_\rho$ = 1.66054 is a unit converter to gram per cubic centimeter | $M_\rho \frac{M}{V}$ |

Since lattice parameter, atomic radius, and atomic volume can be calculated from each other, we use atomic volume and its derivatives (atomic volume mismatch and distortion) in Feature selection.

**Validation of ROM for elastic properties**
Many features such as melting temperature ($T_m$), bulk ($B$), Young's ($Y$), and shear ($\tau$) moduli were estimated from a rule of mixtures (ROM) of the experimental data of constituent elements. The accuracy of the ROM for $T_m$ was evaluated in Refs. [24, 49]. The relative mean absolute error (MAE) for selected RCCAs is about 6%.

$$\begin{cases} \tau_V = \frac{c_{11} - c_{12} + 3c_{44}}{5} \\ \tau_R = \frac{5(c_{11} - c_{12})c_{44}}{4c_{44} + 3(c_{11} - c_{12})} \\ \tau = \frac{\tau_V + \tau_R}{2} \\ Y = \frac{9B\tau}{3B + \tau} \end{cases} \quad (5S)$$

As for elastic constants, DFT calculations provide an error of smaller than 15% for most inorganic phases [48]. Following the procedure proposed by Mehl *et al*. [61], the elastic constants of $c_{11}$, $c_{12}$, and $c_{44}$ of BCC phase of all ternary compositions were calculated. The polycrystalline elastic moduli (Young's and shear moduli) were obtained from the calculated cubic elastic constants by Eqs. (5S, where $\tau_V$ is the Voigt bound [62] for shear modulus and $\tau_R$ is the Reuss bound [63] for shear modulus. The final shear and Young's modulii for the polycrystalline were the Hill [64] average of the Voigt and Reuss bounds, as shown in last two equations in Eqs. (5S).

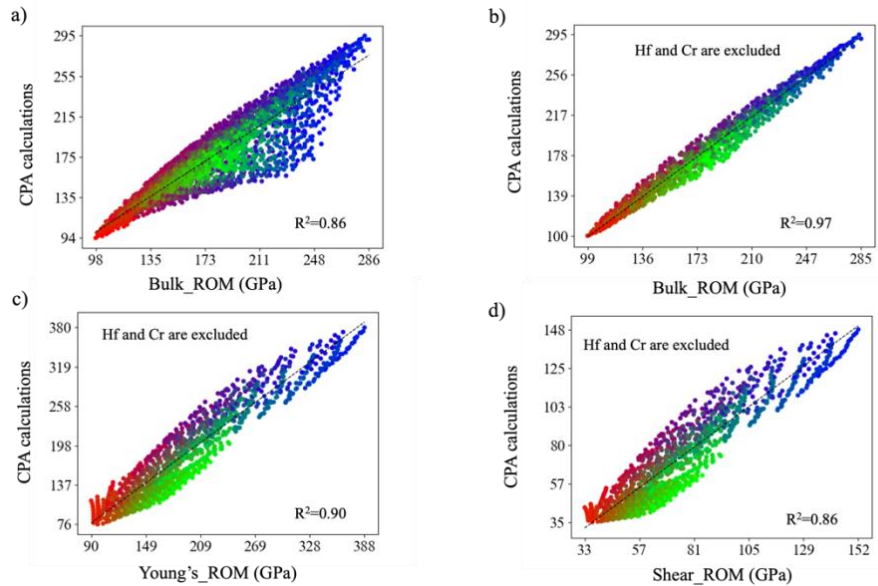

Fig. 9S. Comparison of elastic moduli obtained from CPA calculations and from a rule of mixtures (Bulk_ROM) for ternary compositions: a) bulk modulus for all 3024 ternary compositions; b) bulk modulus for ternary compositions without Hf and Cr elements; c) Young's modulus for ternary compositions without Hf and Cr elements; and d) Shear modulus for ternary compositions without Hf and Cr elements. The datapoints are colored by a normalized RGB (red, green, and blue) color scheme, in which the percentage of the red (red%) is the summed concentrations of Group 4 elements (Ti, Zr, and Hf), green% is the summed concentrations of Group 5 elements (V, Nb, and Ta), and blue% is the summed concentrations of Group 6 elements (Cr, Mo, and W).

The parity plots between the CPA calculated results and ROM results for bulk, Young's, and shear moduli are shown in Fig. 9S. Comparison of elastic moduli obtained from CPA calculations and from a rule of mixtures (Bulk_ROM) for ternary compositions: a) bulk modulus for all 3024 ternary compositions; b) bulk modulus for ternary compositions without Hf and Cr elements; c) Young's modulus for ternary compositions without Hf and Cr elements; and d) Shear modulus for ternary compositions without Hf and Cr elements. The datapoints are colored by a normalized RGB (red, green, and blue) color scheme, in which the percentage of the red (red%) is the summed concentrations of Group 4 elements (Ti, Zr, and Hf), green% is the summed concentrations of Group 5 elements (V, Nb, and Ta), and blue% is the summed concentrations of Group 6 elements (Cr, Mo, and W). For Cr metal, it is reported that DFT method has a problem to fully capture the disordered local moment (DLM) [58], and thus the bulk modulus from CPA calculations produces an error of 37.34% (Table 1S). The big discrepancy in bulk modulus for Hf metal (43.11% in Table 1S) may originate from different valence electron configuration in CPA calculations. By excluding Cr and Hf in these 3024 ternary compositions, the coefficient of determine (R_squared value) of the linear regression (the dashed line) for bulk modulus were improved to 0.97 from 0.86. The R_squared values for Young's and shear moduli were 0.90 and 0.86, respectively. As the elastic moduli that obtained from ROM have a comparable accuracy with CPA calculations, we selected the ROM for the elastic moduli to avoid the relatively large error from DFT calculations in some compositions such as compositions have significant amount of Hf and Cr.

**SHAP values of selected features in Yield strength model**
As important of features might be misleading when strong correlated features are presented in ML models. We thus have computed coefficient of correlation ($R^2$) of all possible combinations of feature pairs in our yield strength model. The high correlated feature pairs ($R^2 > 0.7$) are shown in Fig. 10S. Intent to understand the importance of features that are lower ranked in Fig. 2c in manuscript, we performed SHAP analysis

(Fig. 10S) of $\Delta E_{b0}$, $Y$, $d\tau$, and $dV$ using the room temperature yield strength data. Comparing Fig. 10S with Fig. 2c in manuscript, we believe that $\sigma_{y0}$ and $\Delta E_{b0}$ have similar impacts on predicted yield strength, and likewise with $\chi$ and $Y$.

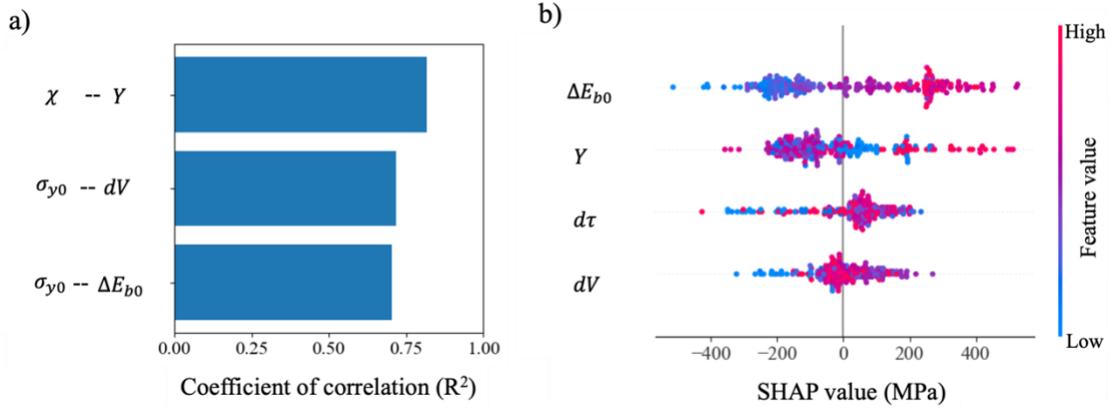

Fig. 10S. Coefficient of correlation ($R^2$) of feature pairs in yield strength model (a), where only highly correlated pair ($R^2 > 0.7$) is shown; b) SHAP value of selected features in yield strength model, where only room temperature yield strength data is used for computing SHAP values.

**Vickers hardness model**
The experimental Vickers hardness ($H_V$ in GPa) were collected and compiled from literatures [23, 37]. Excluding the materials without BCC phase presented in their microstructure, there are 263 datapoints for developing Vickers hardness model. The Tabor model [65-67], in which the hardness is nearly 3 times of the yield strength Eq. (6S), is used for pre-evaluation selection in feature selection procedure in Methods Section in manuscript. The $\sigma_{y0}$ in MC model is selected prior the following feature evaluation process since it was the estimated yield strength at zero-temperature. Following the feature selection process, the final selected features are $\sigma_{y0}$, $d\tau$ (shear modulus distortion), $Y$ (Young's modulus), $\chi$ (electronegativity), and $dV$ (atomic volume distortion). We have developed 20 GBR Vickers hardness model and taken the Vickers hardness model with the highest R_squared value as our final model that used to predict the Vickers hardness for our exhaustive composition space.

$$H_V = 3\sigma_y \tag{6S}$$

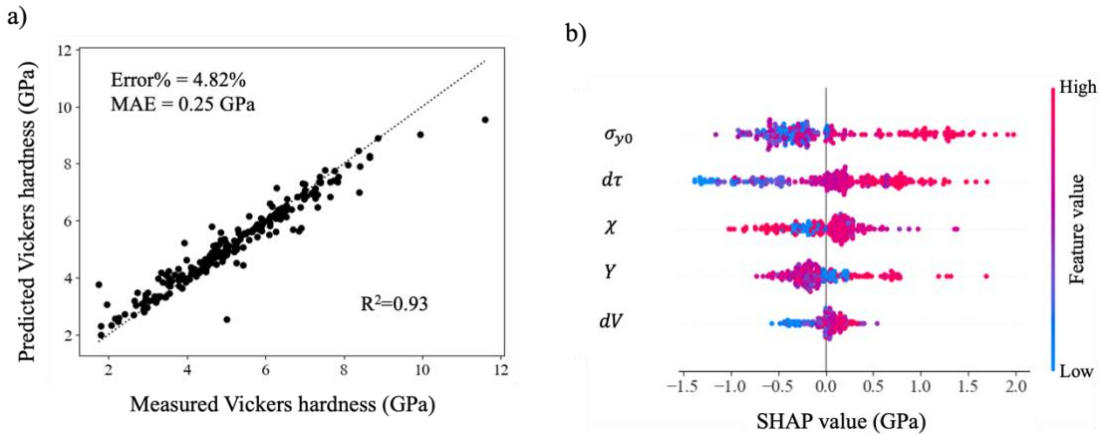

Fig. 11S. Vickers hardness model performance (a) and SHAP value of model features (b).

As shown in Fig. 11S, R_squared value, MAE and error percentage of our Vickers hardness model are 0.93, 0.25 GPa, and 4.82%, respectively. SHAP values presented in Fig. 11S indicated that the Vickers hardness tend to have a positive response to the selected features. The observations are consistent with Tabor model and other theoretic and ML models that summarized in Ref. [22, 23].

**Compression at fracture model**

The experimental compression at fracture ($\varepsilon_f$ in %) were collected from Ref. [37]. If only plastic compression ($\varepsilon_{plas}$) was reported, $\varepsilon_f$ was computed as $\varepsilon_{plas}+\sigma_y/Y$ if both yield strength ($\sigma_y$) and Youngs modulus ($Y$) were reported. There are 150 datapoints collected. Following the feature selection procedure, the selected features for compression model are $d\tau$ (shear modulus distortion), $\nu$ (Poisson's ratio), $B$ (bulk modulus), $\Delta G^{b2h}$ (phase transformation energy from BCC to HCP phases), $\tau$ (shear modulus), and $dV$ (atomic volume distortion). The minimum samples of a leaf node are set to four to prevent from the overfitting and warm start is set to True. We developed 20 Compression at fracture models and taken the model with the highest R_squared value as our final model that used to predict the $\varepsilon_f$ for our exhaustive composition space.

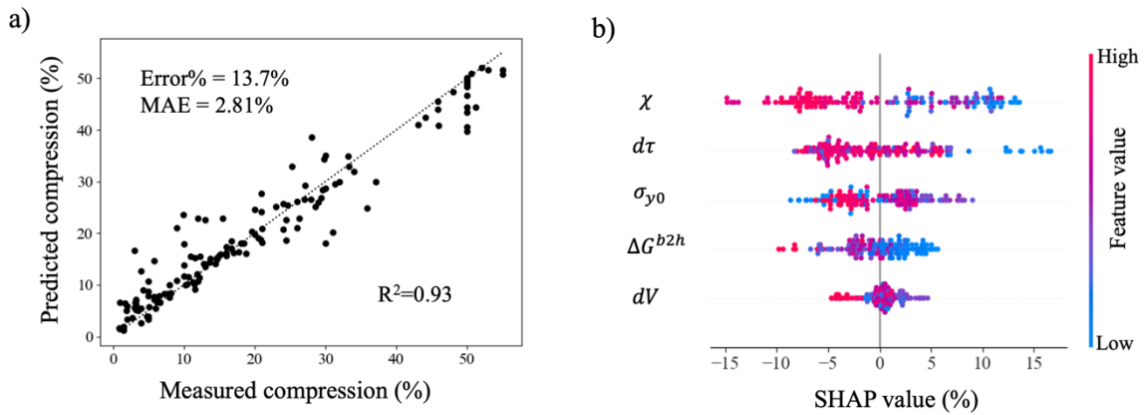

Fig. 12S. Compression model performance (a) and SHAP value of model features (b).

The model performance and SHAP value of model features are shown in Fig. 12S. Our compression model was found a $R^2$ of 0.93, mean absolute error (MAE) of 2.81%, and error percentage of 13.7%.